\documentclass[a4paper]{jpconf}
\usepackage{graphicx}

\newcommand{\be}{\begin{eqnarray}}
\newcommand{\ee}{\end{eqnarray}}
\newcommand{\no}{\nonumber}
\newcommand{\chil}{\chi_{\rm l}}
\newcommand{\chinl}{\chi_{\rm nl}}
\newcommand{\chisg}{\chi_{\rm sg}}

\begin{document}
\title{Energy-gap analysis of quantum spin-glass transitions 
at zero temperature}

\author{Kazutaka Takahashi and Yoshiki Matsuda}

\address{Department of Physics, 
Tokyo Institute of Technology, Tokyo 152--8551, Japan}


\begin{abstract}
 We study for random quantum spin systems 
 the energy gap between the ground and first excited states 
 to clarify a relation to the spin-glass--paramagnetic phase transition.
 We find that for the transverse Sherrington-Kirkpatrick model
 the vanishing of the averaged gap does not identify the transition. 
 A power-law form of the gap distribution function 
 leads to power-law distributions of 
 the linear, spin-glass, and nonlinear susceptibilities.
 They diverge at different points, which we attribute  
 to a quantum ``Griffiths'' mechanism. 
 On the other hand, the classical treatment is justified 
 for the transverse random energy model and 
 the phase transition can be found by 
 a sudden change of the ground state.
\end{abstract}

\section{Introduction}
\label{intro}

 Spin glasses are known as nontrivial states of random spin systems 
 and a lot of studies have revealed their 
 interesting  properties \cite{MPV, FH, Nishimori}.
 Now, the mean field theory of the classical spin-glass 
 has been established and is recognized as a useful tool 
 for attacking various random systems.
 As one of such applications, 
 quantum systems have been studied for many years 
 since the work by Bray and Moore \cite{BM}.
 Recently, much attention has been paid to quantum systems 
 for their applications to information processing 
 such as quantum annealing \cite{FGSSD, KN, FGGLLP}.
 In contrast to classical systems,  
 mean field analysis has a serious limitation
 and a reliable analytical method is needed.

 At finite temperature, 
 state of the system is determined by the free energy.
 A competition between energy and entropy 
 induces a phase transition between different phases. 
 At zero temperature, the system is determined by the energy only 
 and the entropy does not play any role.
 The quantum fluctuations become important at low temperature  
 and the system falls into its ground state.
 Then, the phase transition is realized 
 by a change of the ground state, which can be found by looking at  
 the energy gap between the ground and first excited states.
 Although the mechanism of a quantum phase transition 
 is simpler than that of a thermodynamic phase transition, 
 it is rather difficult to treat quantum fluctuation effects
 even for clean systems \cite{Sachdev}.
 The effect of disorder further makes 
 the analysis even more difficult.
 As in the classical system, the mean field theory of 
 quantum spin-glass systems has been developed.
 However, a lot of studies use the static approximation \cite{BM}
 which neglects the quantum fluctuation effects.
 Although this approximation is justified for not-too-low temperatures, 
 no reliable result can be obtained at zero temperature. 

 In the present paper,
 we give a detailed study of quantum phase transitions 
 in terms of the energy gap \cite{TM}.
 As typical random quantum systems 
 which exhibit a spin-glass phase transition, 
 we treat the Sherrington-Kirkpatrick (SK) model \cite{SK} and 
 random energy model (REM) \cite{Derrida} with a transverse field.
 The transverse REM can be solved exactly and 
 shows a discontinuous transition \cite{Goldschmidt}.
 The transverse SK model has not been solved exactly, but 
 many studies indicate a continuous transition.
 Since these models without transverse field 
 can be treated exactly by a mean field theory,
 most of previous studies took a similar approach.
 We study these systems by using the energy gap analysis.
 In section \ref{level}, 
 we define the models and consider the energy gap 
 by using perturbation theory.
 The energy gap is also examined numerically and 
 we see that the gap strongly fluctuates from sample to sample. 
 In order to see a phase transition, several kinds of susceptibilities 
 are studied in terms of the energy gap in section \ref{sus}.
 We find there a strong quantum fluctuation effect.
 After discussing a sufficient condition for this effect to occur 
 in section \ref{ipr}, 
 we discuss its mechanism in section \ref{qgs}. 
 The final section \ref{conc} is devoted to give 
 conclusions, perspectives, and possible applications.

\section{Energy level distribution}
\label{level}

 In this section, we define quantum spin-glass models and 
 examine the energy gap between the ground and first excited states.

\subsection{Model}

 We treat the $p$-body interacting spin-glass model 
 with transverse field 
\be
 \hat{H} = -\sum_{i_1<i_2<\cdots<i_p}^{N}J_{i_1i_2\cdots i_p}
 \sigma^z_{i_1}\sigma^z_{i_2}\cdots\sigma^z_{i_p}
 -\Gamma\sum_{i=1}^N\sigma^x_i, \label{GSK}
\ee
 where $\sigma_{i}^{z,x}$ are Pauli matrices on site $i$, 
 $N$ is the number of the site, and $\Gamma$ is the transverse field.
 We take an ensemble average over random interaction 
 $J_{i_1i_2\cdots i_p}$ with a probability distribution 
\be
 {\rm P}(J_{i_1\cdots i_p})=
 \sqrt{\frac{N^{p-1}}{\pi p! J^2}}
 \exp\left\{
 -\frac{N^{p-1}}{p!J^2}
 \left(J_{i_1i_2\cdots i_p}\right)^2
 \right\},
\ee
 where $J$ is the strength of the random fluctuations.
 Here we study the cases of $p=2$ (SK model) and 
 $p\to\infty$ (REM),  
 and the case of other-$p$ will be reported elsewhere.
 The quantum REM can be solved exactly 
 and at zero temperature
 the system shows a discontinuous phase transition between 
 the spin-glass and paramagnetic phases at 
 $\Gamma/J=\sqrt{\ln 2}\sim 0.83$ \cite{Goldschmidt}.
 The SK model has not been solved exactly, 
 but a lot of studies reported a continuous transition 
 around the point $\Gamma/J = 1\sim 2$ 
 when the temperature is equal to zero.
 A naive perturbative calculation gives $\Gamma/J=1$ 
 and the static approximation 
 gives $\Gamma/J=2$ \cite{Usadel, TLK}, 
 while more advanced methods indicate a point around 
 $\Gamma/J\sim 1.5$ \cite{YI, GL, MH, AR, ADDR, Takahashi}.

\subsection{Perturbative calculation of the energy gap}

 Here we estimate for the SK model the energy gap between 
 the ground and first excited states by using 
 a perturbative expansion with respect to $J/\Gamma$.
 When the randomness $J$ is absent, the Hamiltonian is exactly solved
 to give energy levels $E_n^{(0)}=-(N-2n)\Gamma$ 
 with nonnegative integer $n$.
 Each of the levels has a degeneracy $N!/n!(N-n)!$ 
 which is split into many nondegenerate ones for finite-$J$.
 The unperturbed ground state is denoted by $n=0$
 and the first excited state belongs to the sector with $n=1$.
 The Hamiltonian of the $n=1$ sector is given by 
 $H^{(1)}_{ij}=-(N-2)\Gamma-J_{ij}$.
 Since $J_{ij}$ are Gaussian random variables, 
 the energy level distribution is given by 
 that of the Gaussian orthogonal ensemble \cite{Mehta}.  
 The energy levels distribute in a semicircle form 
 and the lowest energy level is given by $E_1=-(N-2)\Gamma-2J$.
 Neglecting the correlations between blocks with different $n$'s, 
 we obtain the energy gap $\Delta\sim 2(\Gamma-J)$
 and the degenerate point $\Gamma=J$.

 The subleading corrections are calculated as follows.
 The ground state $n=0$ is coupled to $N(N-1)/2$-levels in 
 the sector $n=2$ by random interactions $J_{ij}$.
 Up to second order in perturbation theory, 
 we obtain in average,  
\be
 E_0^{(2)} \sim -N\Gamma 
 -\frac{\sum_{i<j}^{N}J_{ij}^2}{E_2^{(0)}-E_0^{(0)}}
 \sim -N\Gamma-\frac{J^2}{8\Gamma}(N-1).
\ee
 In the same way, the first excited state is directly coupled to 
 the sector $n=3$.
 The correction is roughly estimated as 
\be
 E_1^{(2)} \sim -(N-2)\Gamma -2J 
 -\frac{J^2}{8\Gamma}(N-3).
\ee
 As a result we have the energy gap
\be
 \Delta\sim 2(\Gamma-J)+\frac{J^2}{4\Gamma}, 
\ee
 and the degenerate point $\Gamma/J= (2+\sqrt{2})/4\sim 0.85$.
 Although the present analysis is based on a perturbative expansion 
 and the degenerate point cannot be estimated exactly, 
 it is probable to say that the averaged degenerate point 
 does not coincide with the phase transition point around 1.5.
 On the other hand, a similar perturbative calculation was carried out
 for the transverse REM \cite{JKKM} and 
 a good estimate of the phase transition point was obtained.

\subsection{Numerical result of the energy gap}

 We estimate the energy gap by using the numerical calculation.
 The Hamiltonian of the size $2^N$ is diagonalized 
 by using the Lanczos method and 
 the average is taken over more than 20000 samples.
 The number of the site is taken to be $N=8-22$.
 We note that for the SK model, 
 the Hamiltonian commutes with $P=\prod_{i=1}^N\sigma_i^x$
 and the matrix splits into two blocks.
 This allows us to diagonalize their two matrices separately.
 For the REM, we take the Derrida's original Hamiltonian 
 for the random part and do not use the original Hamiltonian
 (\ref{GSK}) with $p\to\infty$. 
 The Gaussian random variables with the average 
 $\left[E_i\right]=0$ and the variance 
 $\left[E_i^2\right]=NJ^2/2$ are produced $2^N$ times 
 and are used for the diagonal part of the Hamiltonian in $\sigma_z$-basis.

\begin{figure}[th]
\begin{minipage}{18pc}
\includegraphics[width=18pc]{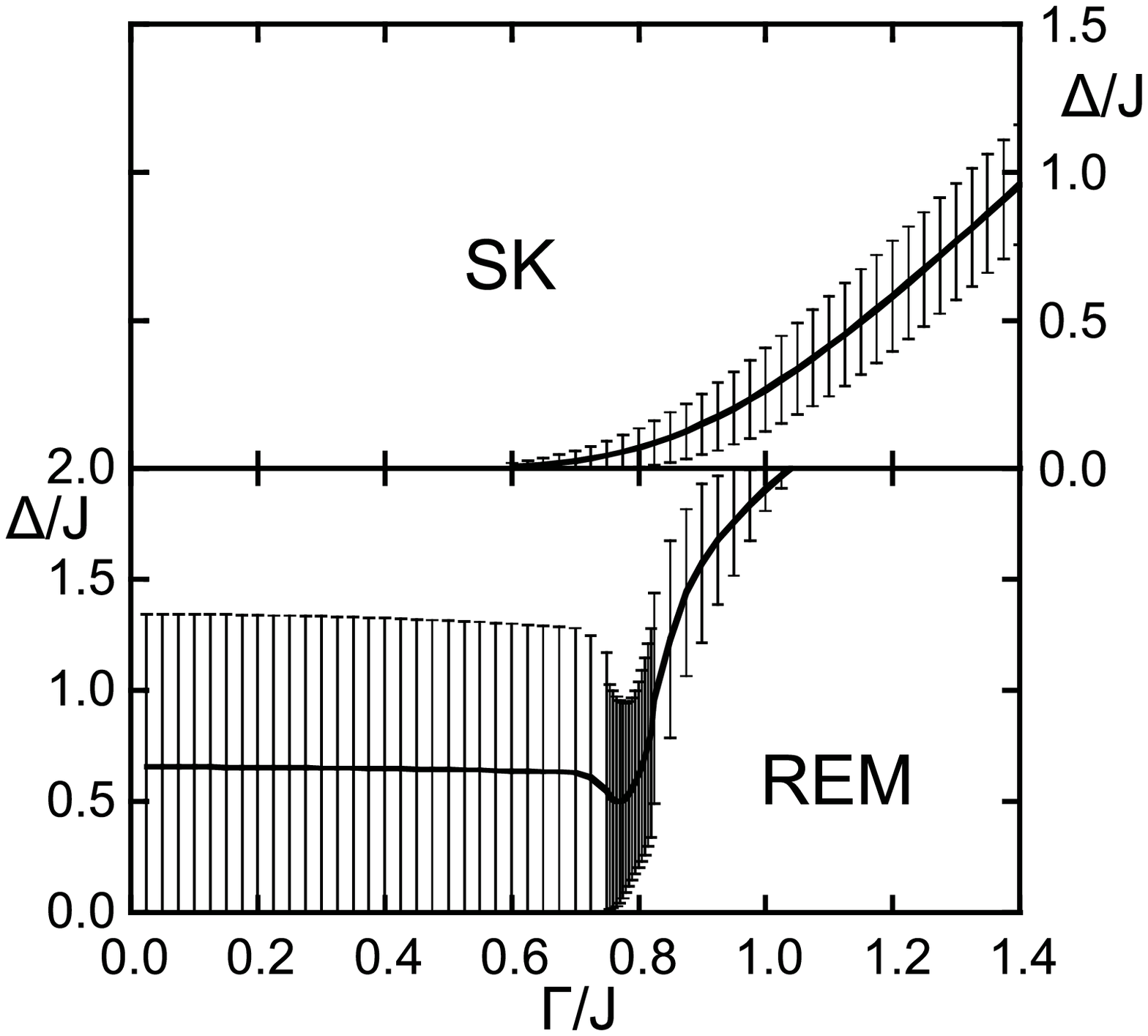}
\caption{\label{gap}Averaged energy gap $\Delta=[E_1-E_0]$
 between the ground and first excited states
 for the SK model and REM with $N=22$.
 The variance of the energy gap $\pm\sqrt{[(E_1-E_0)^2]-\Delta^2}$ 
 is denoted by the error bar. }
\end{minipage}\hspace{2pc}%
\begin{minipage}{18pc}
\includegraphics[width=18pc]{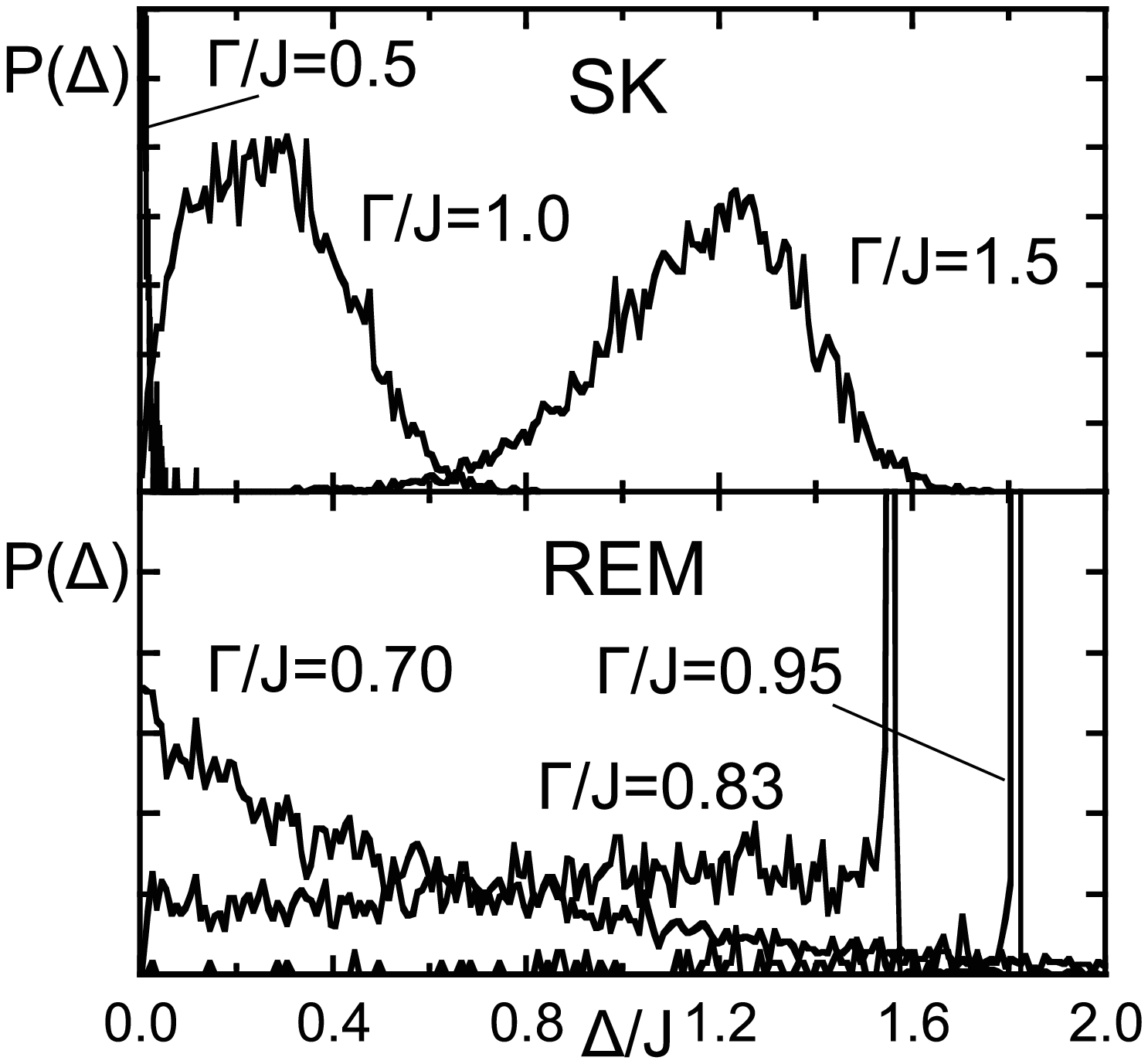}
\caption{\label{pg}Gap distribution function
 for the SK model and REM with $N=22$.
 For the REM, 
 $\Gamma/J\sim 0.83$ corresponds to the phase transition point.}
\end{minipage}
\end{figure}

 In figure \ref{gap}, we plot the averaged energy gap.
 For the SK model, the gap is an increasing function of $\Gamma$
 and becomes negligible at values smaller than $\Gamma/J \sim  0.6$.
 This result is consistent with the perturbative calculation.
 From this figure, we see that it is hard to identify 
 a phase transition point around $\Gamma/J\sim 1.5$ 
 even if the fluctuations are taken into account.
 On the other hand, the gap shows a complicated behavior for the REM.
 The averaged gap becomes minimum around the phase transition point 
 and the fluctuations behave differently on both sides of the point.
 These results imply that it is possible to find a phase transition 
 from the averaged energy gap for the REM, 
 while it is hard for the SK model.

 In order to see the difference more in detail, 
 we show the energy gap distribution function in figure \ref{pg}.
 For the SK model, we find long tails of the distribution function.
 It is hard to imagine that these fluctuations vanish 
 at the thermodynamic limit and
 we expect a significant role of fluctuations.
 A more complicated behavior 
 can be observed for the REM.
 A single peak is replaced by another one, 
 which implies a discontinuous transition between two different states.

 These analytical and numerical results show that 
 for the SK model the fluctuations play a significant role 
 and the averaged gap is not a useful quantity 
 to describe the phase transition. 
 Concerning the REM, the averaged gap is expected to be useful
 although the fluctuations are large as well. 
 Therefore, it is important to understand 
 where this difference comes from.

\section{Susceptibility}
\label{sus}

 In the standard theory of critical phenomena, 
 phase transitions are identified by a singularity of 
 response functions such as the magnetic susceptibility.
 In this section, 
 we discuss the linear, spin-glass, and nonlinear susceptibilities.
 For our purpose to find a transition in terms of the energy gap, 
 the spectral representation is useful and is discussed in detail.

\subsection{Spectral representation}

 The magnetization of the system is given by the response of 
 the longitudinal magnetic field $h$ 
 whose contribution to the Hamiltonian 
 is given as $\hat{H}\to\hat{H}-h\sum_i\sigma_i^z$.
 We have 
\be
 m = \frac{1}{N\beta}\frac{\partial}{\partial h}
 \left[\ln Z\right], 
\ee
 where $Z=\Tr e^{-\beta \hat{H}}$ is the partition function and 
 $\beta=1/T$ is the inverse temperature.
 The magnetization can be expanded with respect to $h$
 as $m=\chi_{\rm l}h-\chi_{\rm nl}h^3+\cdots$, 
 which defines the linear ($\chi_{\rm l}$) and 
 nonlinear ($\chi_{\rm nl}$) susceptibilities \cite{Chalupa, Suzuki}.

 Our analysis is based on the imaginary time formalism
 \cite{Sachdev, NO}.
 The linear susceptibility is expressed as 
\be
 \chil = \frac{2}{N\beta}\sum_{i,j=1}^N
 \int_0^\beta d\tau\int_0^\tau d\tau'
 \left[\frac{1}{Z}
 \Tr e^{-\beta\hat{H}}\sigma_i^z(\tau)\sigma_j^z(\tau')\right]
 -\beta q, 
 \label{chiltau}
\ee
 where $\sigma_i^z(\tau)=e^{\hat{H}\tau}\sigma_i^z e^{-\hat{H}\tau}$, 
 $\tau$ is the imaginary time running from 0 to $\beta$, and 
 $q$ is the spin-glass order parameter.
 Inserting the completeness relation of eigenstates of
 the Hamiltonian, we obtain that the first term of 
 the right hand side of equation (\ref{chiltau}) is equal to 
\be
 \frac{\beta}{N}\left[\frac{1}{Z}
 \sum_{mn (E_m=E_n)}e^{-\beta E_n}
 |\langle n|\sum_i\sigma_i^z|m\rangle|^2\right] 
 +\frac{2}{N}\left[
 \frac{1}{Z}\sum_{mn (E_m\ne E_n)}\frac{e^{-\beta E_n}}{E_m-E_n}
 |\langle n|\sum_i\sigma_i^z|m\rangle|^2\right],
 \label{chil1}
\ee
 where $|n\rangle$ is an eigenstate with the energy $E_n$.
 The first term comes from degenerate states
 and the second from nondegenerate ones.
 The first term is equal to the static part $\omega=0$ 
 of the spectral function 
\be
 \chi(\omega) = 
 \frac{1}{N}\left[
 \frac{1}{Z}\sum_{nm}\delta(\omega-E_m+E_n)
 e^{-\beta E_n}
 |\langle n|\sum_i\sigma_i^z|m\rangle|^2\right].
 \label{chiw}
\ee 
 For the static part $\chi=\chi(\omega=0)$, 
 the energy is conserved through a virtual process in imaginary time
 and a classical treatment is possible.
 Then, the linear susceptibility is expressed by a sum of
 classical and quantum parts.
 The classical part $\chil^{({\rm c})}$ is equal to $\beta(\chi-q)$
 and can be treated by the static approximation.
 It corresponds to the linear susceptibility 
 for the classical system given by $\chil=\beta(1-q)$ \cite{FH}.
 On the other hand, the quantum part is expressed as 
 the second term of equation (\ref{chil1}).
 At the zero temperature limit, we assume that the ground state 
 $|0\rangle$ with the energy $E_0$ is nondegenerate and 
 obtain for the quantum part 
\be
 \chil^{({\rm q})} = \frac{2}{N}\left[\sum_{n\ne 0}
 \frac{\left|\langle 0 |\sum_{i}\sigma_i^z|n\rangle\right|^2}{E_n-E_0}
 \right]. 
\ee
 In the same way, the quantum part of 
 the nonlinear susceptibility is given by
\be
 \chinl^{(\rm q)}
 &=& \frac{4}{N}\left[\sum_{n,m\ne 0}
 \frac{|\langle 0|\sum_{i}\sigma_i^z|n\rangle|^2}{E_n-E_0}
 \frac{|\langle 0|\sum_{j}\sigma_j^z|m\rangle|^2}
 {(E_{m}-E_0)^2}\right]
 \no\\ & & 
 -\frac{4}{N}\left[\sum_{n,n',n''\ne 0}
 \frac{ \langle 0|\sum_{i}\sigma_i^z|n\rangle
 \langle n|\sum_{j}\sigma_j^z|n'\rangle
 \langle n'|\sum_{k}\sigma_k^z|n''\rangle
 \langle n''|\sum_{l}\sigma_l^z|0\rangle}
 {(E_n-E_0)(E_{n'}-E_0)(E_{n''}-E_0)}\right].
 \label{chinl}
\ee

 In classical spin-glass systems, 
 the nonlinear susceptibility is an important quantity
 since it is related directly to the spin-glass susceptibility 
 $\chisg$ as $\chinl=\beta(\chisg-2\beta^2/3)$. 
 A spin-glass transition
 which is characterized by the divergence of $\chisg$ 
 can be observed directly by $\chinl$.
 However, in the present case, 
 the quantum part of the spin-glass susceptibility is given by 
\be
 \chisg^{(\rm q)}=
 \frac{4}{N}\left[\left(\sum_{n\ne 0}
 \frac{\left|\langle 0 |\sum_i\sigma_i^z|n\rangle\right|^2}{E_n-E_0}
 \right)^2\right].
\ee
 $\chinl^{({\rm q})}$ and $\chisg^{({\rm q})}$ behave 
 differently and there is no direct relation between them. 
 The spin-glass susceptibility is not equivalent to 
 the nonlinear susceptibility, which 
 is a crucial difference from the classical system.

 If the transition is of quantum nature, 
 the singularity comes from the quantum part.
 We see from above expressions that 
 the three susceptibilities are divergent when 
 the gap between the ground and first excited states 
 collapses into zero.
 This is understood as the standard mechanism of 
 a quantum phase transition as we mention in section \ref{intro}.
 However, in random systems, 
 the gap vanishing point fluctuates from sample to sample.
 If the fluctuation is small, the transition point can be 
 given by the averaged gap vanishing point.
 But, as we see in section \ref{level},  
 the fluctuation is not negligible 
 and the vanishing point does not coincide with 
 the transition point reported in previous studies.
 If the gap distribution function behaves in a power-law form 
 ${\rm P}(\Delta)\sim \Delta^k$, 
 we can find that three susceptibilities diverge at different points.
 Since $\chil$, $\chisg$, and $\chinl$ are roughly proportional to 
 the gap as $1/\Delta^{1, 2, 3}$ respectively,
 they diverge at $k=0, 1, 2$ respectively.
 We note that this behavior is specific to random and quantum systems.
 The quantum nature is necessary to obtain the spectral representations
 and a power-law distribution of the gap 
 can be obtained only for random systems.

\subsection{Numerical result}

\begin{figure}[th]
\begin{minipage}{18pc}
\includegraphics[width=18pc]{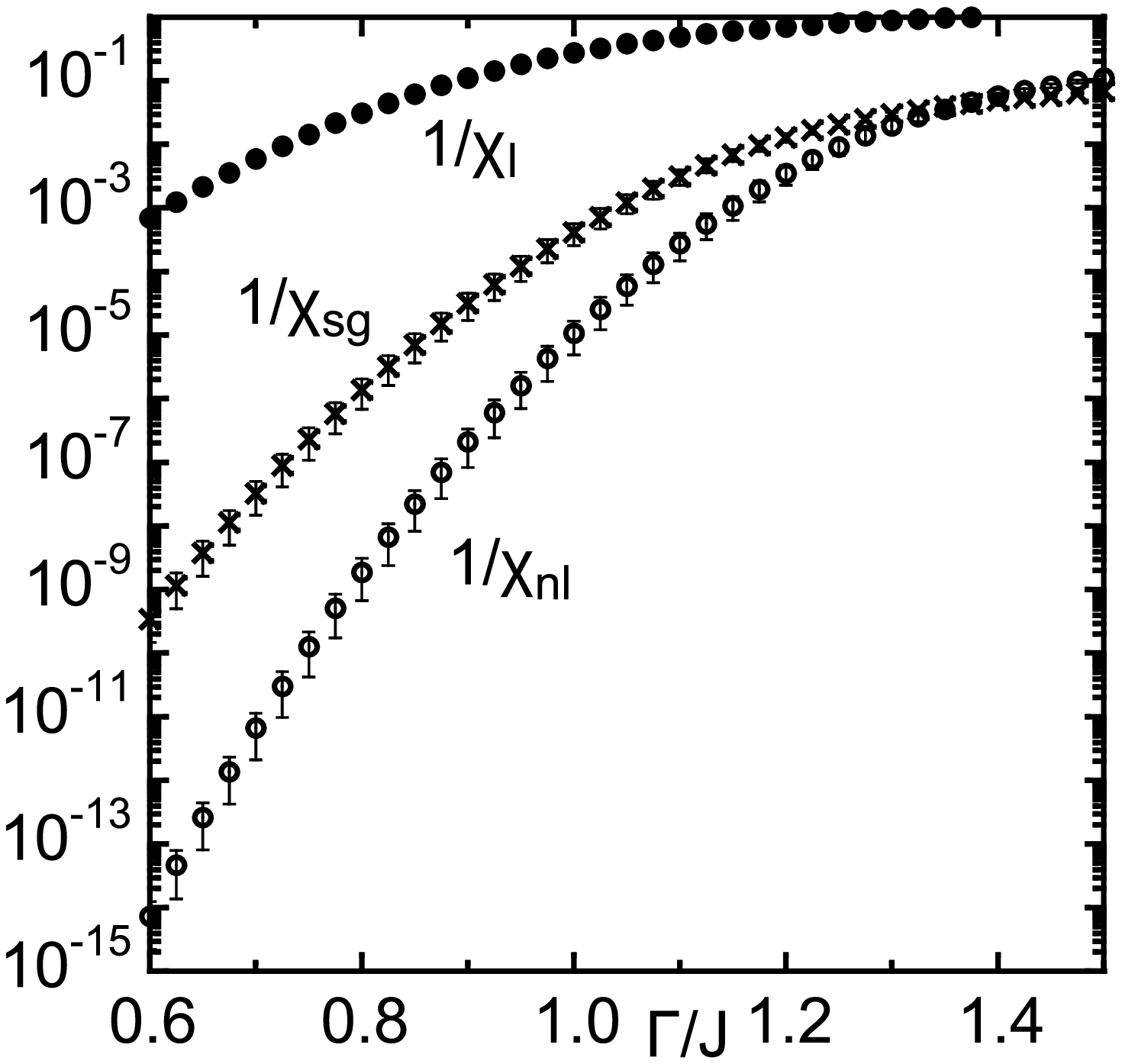}
\caption{Inverse susceptibilities (quantum part)
 for the SK model with $N=18$. 
}
\label{susfig}
\end{minipage}\hspace{2pc}%
\begin{minipage}{18pc}
\includegraphics[width=18pc]{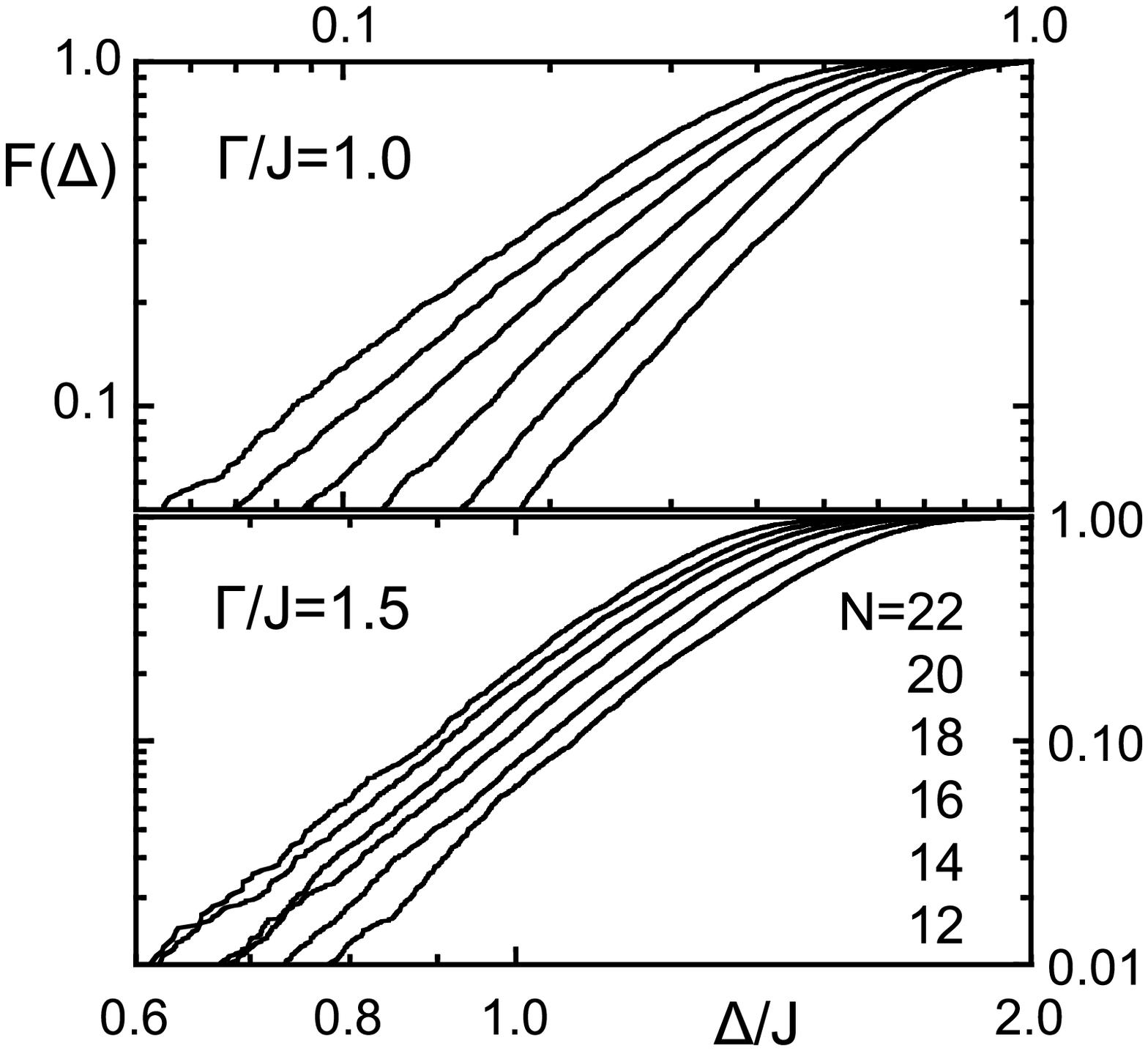}
\caption{The cumulative distribution function of the energy gap 
 $F(\Delta)=\int_0^\Delta d\Delta' P(\Delta')$
 ($N=12$,14,16,18,20, and 22 from below to above)
 for the SK model.
}
\label{cdf}
\end{minipage}
\end{figure}

 In order to see whether the above scenario holds, 
 we analyze numerically 
 the spectral representation of the susceptibilities.
 In figure \ref{susfig}, 
 we plot the quantum part of $\chil$, $\chisg$, and $\chinl$ 
 for the SK model.
 They become very large at small transverse fields 
 and there is a relevant difference 
 between $\chisg^{(q)}$ and $\chinl^{(q)}$, 
 which support our expectation.
 We also plot the cumulative distribution function of 
 the energy gap in figure \ref{cdf}.
 The size dependence of the function implies that 
 the power-law persists in the thermodynamic limit.

\begin{figure}[th]
\begin{minipage}{18pc}
\includegraphics[width=18pc]{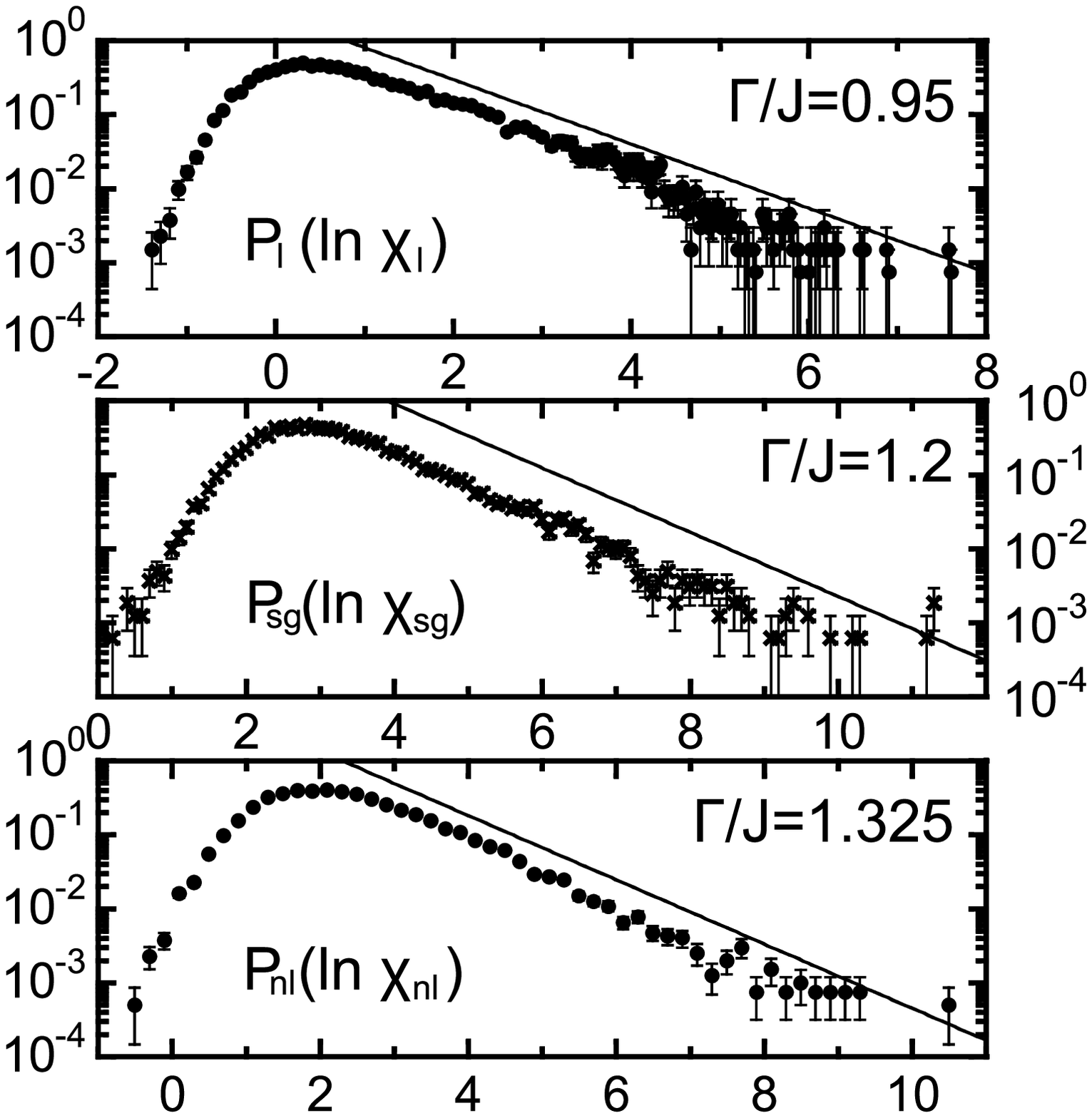}
\caption{Susceptibility distribution functions  
 for the SK model with $N=18$.
 The straight lines are ${\rm P}(\chi)\sim 1/\chi^{2}$.
}
\label{susdist}
\end{minipage}\hspace{2pc}%
\begin{minipage}{18pc}
\includegraphics[width=18pc]{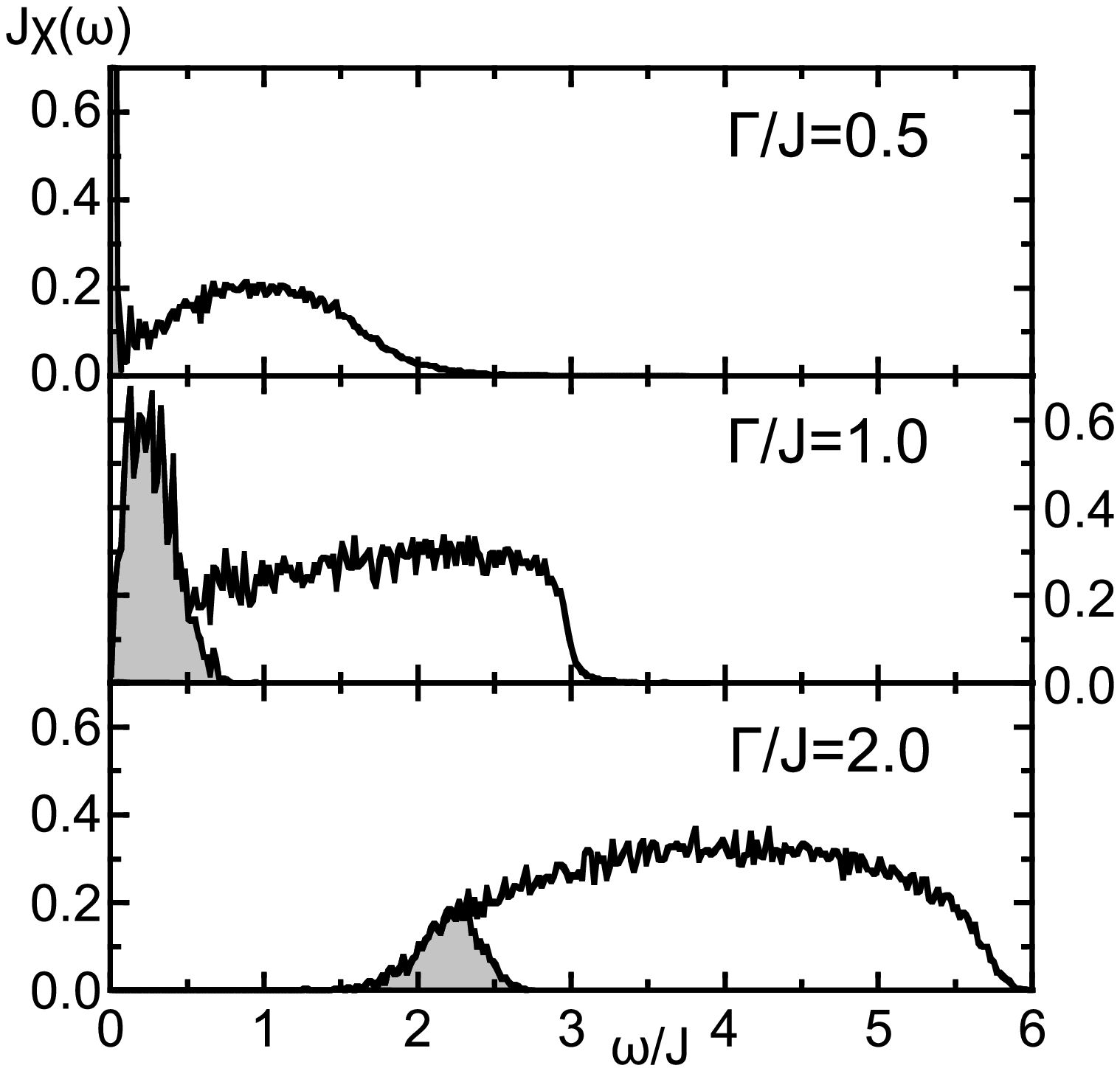}
\caption{Spectral function $\chi(\omega)$ 
 for the SK model ($N=20$).
 The shaded parts are contributions from the first excited state.
}
\label{chi}
\end{minipage}
\end{figure}

 We also plot the distribution function of 
 the susceptibilities in figure \ref{susdist}.
 The distribution function ${\rm P}(\chi)$
 shows a power-law dependence at large-$\chi$ as 
 ${\rm P}(\chi)\sim \chi^{-s-1}$.
 In this form of the distribution function, 
 the averaged susceptibility diverges when $s<1$.
 We analyze the distribution functions  
 for each of the susceptibilities 
 and find that the divergent point can be different with each other,
 which is consistent with the analysis of the gap distribution. 
 Although our system size is not so large 
 enough to give a reliable extrapolation value,   
 our result of the spin-glass transition,   
 identified by the divergence of $\chisg$, 
 supports the results in 
 \cite{YI, GL, MH, AR, ADDR, Takahashi}.
 Previous studies did not pay any attention to 
 the difference between the spin-glass and nonlinear susceptibilities
 and it may be worth reconsidering the results more carefully.

 We note that the present numerical analysis is 
 only for the quantum part.
 To find an observable physical value of susceptibilities, 
 we must include the classical part.
 The classical part of the susceptibilities 
 can be analyzed by the static approximation.
 For the linear susceptibility, $\chil^{(c)}=\beta(\chi-q)$ 
 only gives a cusp and does not show any strong singularity.
 The classical nonlinear and spin-glass susceptibilities 
 can be divergent and we must study closely which part between 
 the classical and quantum parts is important.
 It can be considered that 
 the classical part is important at the spin-glass phase
 and the quantum part is dominant at the paramagnetic phase.
 We treat a related problem in the next section 
 and a detailed study is presented elsewhere. 

\subsection{Effect of higher excited states}

 Up to here, our attention has been paid to 
 the gap between the ground and first excited states.
 The spin-glass system has a complicated ground state 
 which is described by a replica-symmetry-broken order parameter, 
 and it might be natural to expect that at the transition point 
 a vast amount of degeneracy occurs at the same time.
 In order to see effects of higher order excited states, 
 we plot the spectral function (\ref{chiw}) 
 for the SK model in figure \ref{chi}.
 For large-$\Gamma$, 
 the function was analytically calculated as \cite{TT}
\be
 \chi(\omega) = \frac{\sqrt{4J^2-(\omega-2\Gamma)^2}}{2\pi J^2}
 \theta(2J-|\omega-2\Gamma|).
\ee
 In this case, the contribution from the first excited state 
 cannot be distinguished from those of other states. 
 This property drastically changes when $\Gamma$ is small.
 We see that a single peak from the first excited states 
 is separated from other contributions and 
 plays a dominant role for the phase transition.
 Actually, when the transverse field is absent, 
 we have a symmetry that 
 the Hamiltonian is unchanged under the gauge transformation  
 $\sigma_i^z\to -\sigma_i^z$. 
 As a result, the energy level is doubly degenerate 
 in the classical limit.
 Thus, for the present model, 
 the first excited state plays a dominant role 
 for the phase transition.

 This result implies that the second term of the right hand side of
 equation (\ref{chinl}) gives a subleading correction 
 since the expression includes matrix elements of 
 higher order excited states ($|n'\rangle$).
 Numerically we confirm that this term is not important 
 to find a divergent point, 
 which makes numerical calculations easier.

\section{Localization of state}
\label{ipr}

 In the previous section, we have treated the SK model only.
 The spectral representation is a general relation and 
 is independent of the explicit form of the Hamiltonian.
 However, it is evident from the known exact solution that 
 the above scenario cannot be applied to the REM.
 Actually the classical treatment is justified for the REM 
 and the quantum part of the susceptibility does not 
 give any significant contributions.
 This can be understood by examining the inverse participation ratio 
 (IPR) defined as 
\be
 I = \sum_{n}\left|\langle n|\sigma_i^z|0\rangle\right|^4.
 \label{ipreq}
\ee
 This function describe how many states are included 
 in the state $\sigma_i^z|0\rangle=\sum_{n=1}^M c_n|n\rangle$.
 If it includes $M$-states, 
 $c_n\sim 1/\sqrt{M}$ and we obtain $I\sim 1/M$.
 Thus this function is a useful quantity to see 
 whether the state is localized in eigenstate space \cite{Wegner}.

\begin{figure}[th]
\begin{minipage}{18pc}
\includegraphics[width=18pc]{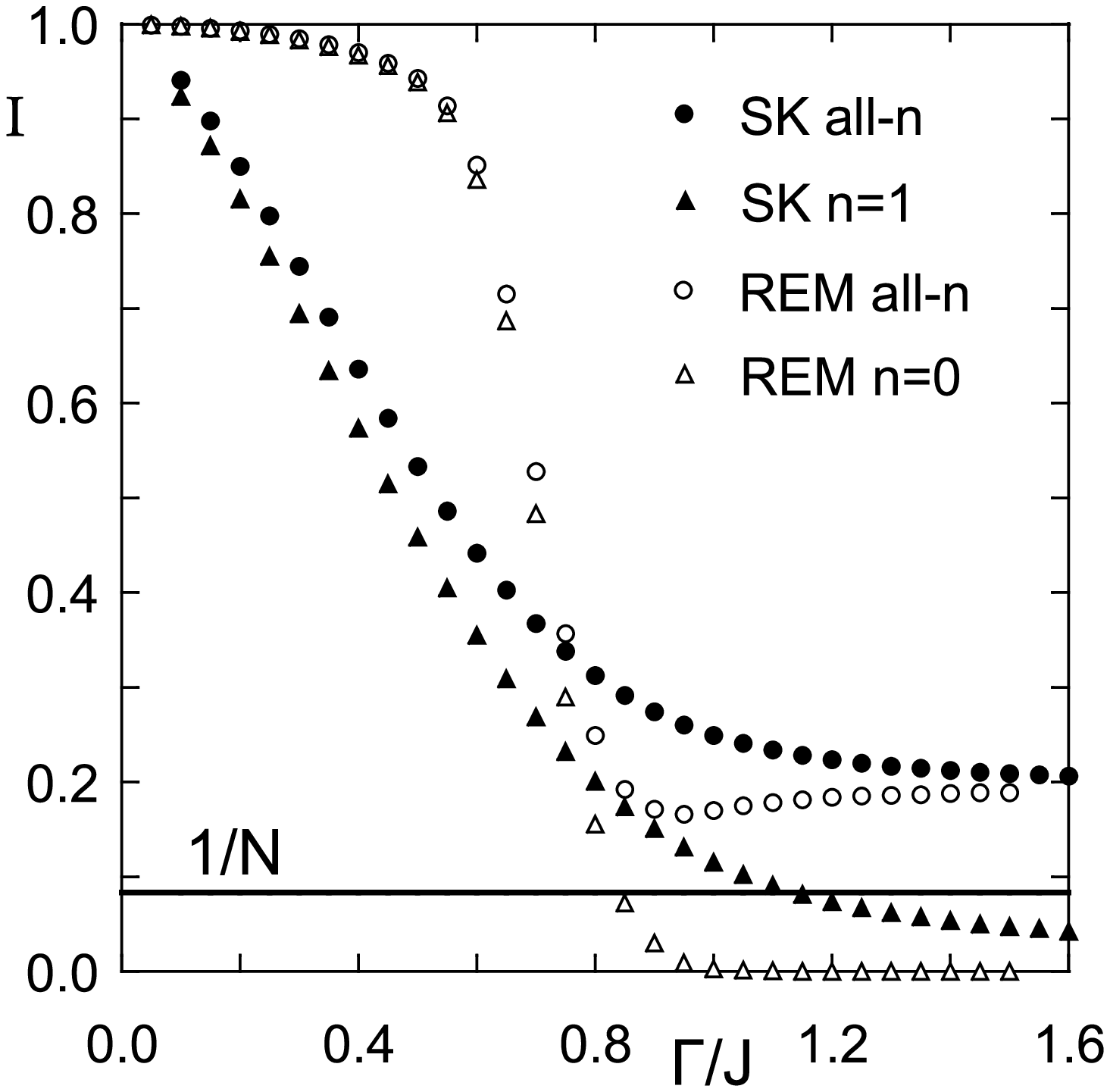}
\caption{Inverse participation ratio (N=12).
 The summation over $n$ in equation (\ref{ipreq}) is restricted to 
 $n=0 (1)$  for a plot with caption $n=0 (1)$.
}
\label{iprfig}
\end{minipage}\hspace{2pc}%
\begin{minipage}{18pc}
\includegraphics[width=18pc]{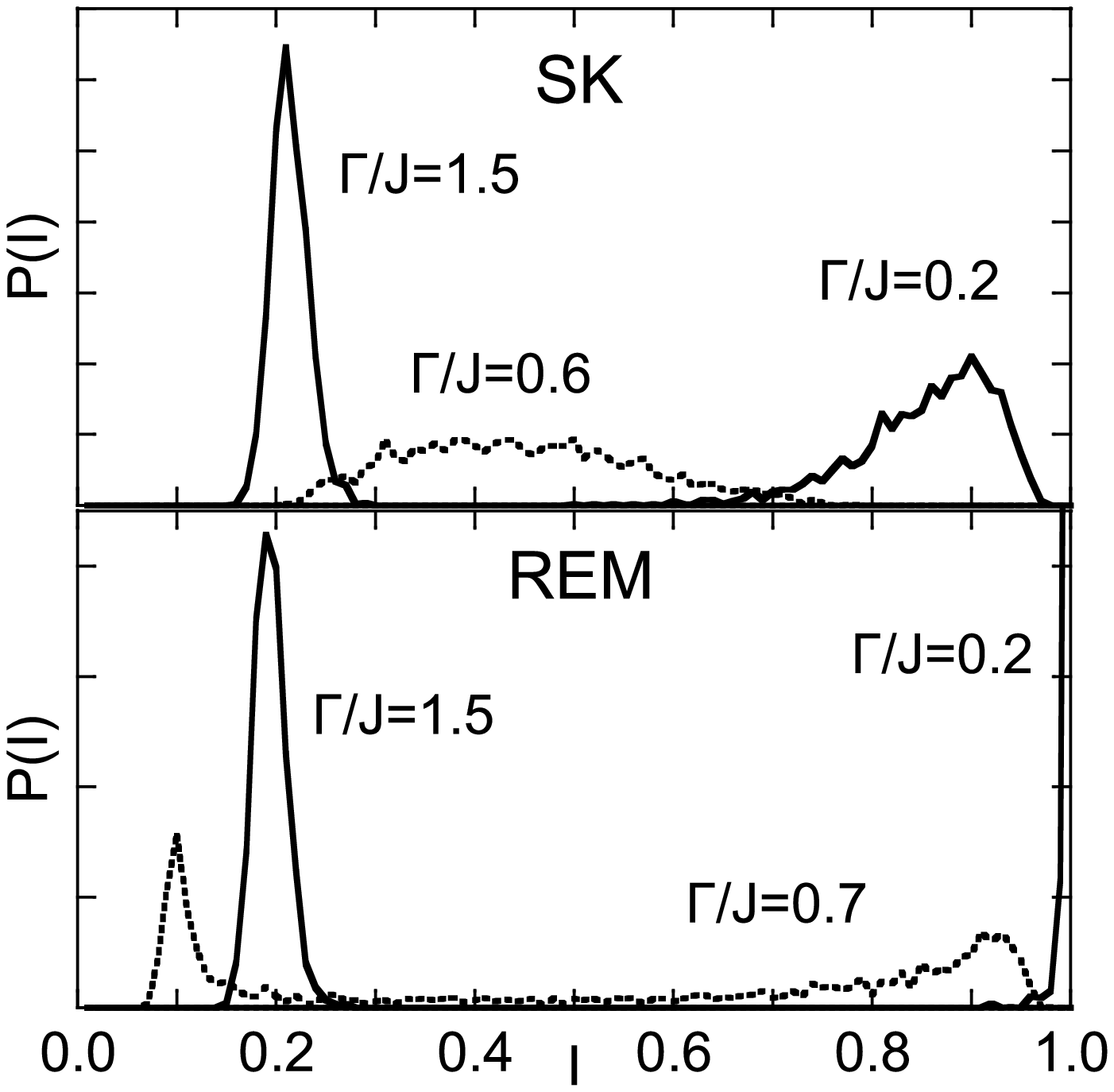}
\caption{Distribution function of the inverse participation ratio (N=12).
}
\label{iprdist}
\end{minipage}
\end{figure}

 In figure \ref{iprfig}, we plot the averaged IPR 
 for both models and find a significant difference between them.
 For the SK model, the IPR decreases uniformly 
 from 1 to $1/N$ as a function of $\Gamma$. 
 This behavior reflects the fact that 
 the gauge symmetry is important at small-$\Gamma$
 and, at large-$\Gamma$, the one spin flip changes the ground state 
 to the sector $n=1$ with $N$-levels.
 $\sigma_i^z|0\rangle$ mainly consists of 
 the first excited state.
 There is no such symmetry for the REM and 
 the IPR becomes minimum around 
 the phase transition point.
 At large-$\Gamma$ the IPR gradually approaches to unity.
 $\sigma_i^z|0\rangle$ almost stays in the ground state $|0\rangle$
 and the excitations to the higher energy levels are small.
 This result shows that the quantum fluctuation effect can be 
 negligible for the REM.
 This is the reason why the discussion in the previous section 
 cannot be applied to the REM.
 The state of the system is determined by the ground state.

 We also find for the IPR distribution function 
 in figure \ref{iprdist} that the IPR is always larger than $1/N$.
 This shows that all states $2^N$ do not contribute 
 to determine the state of the system, 
 which may allow us to construct a reduced effective model.
 As in the energy gap distribution, 
 a double peak structure is observed for the REM,
 while it is not for the SK model.

\section{Quantum Griffiths singularity}
\label{qgs}

 The power-law distribution of the susceptibility 
 was also found in finite-dimensional systems 
 \cite{Fisher1, GBH1, RY1, Fisher2, RY2, GBH2}.
 In two and three dimensional systems,
 the nonlinear susceptibility 
 diverges at a point of $\Gamma$ which 
 is larger than the phase transition point. 
 This was understood the Griffiths singularity 
 effect \cite{Griffiths, McCoy}.
 Rare configurations of ordered clusters in a disordered state
 give a singular contribution to the susceptibility.
 A similar effect is known in disordered electron systems
 as anomalously localized states \cite{AKL}.

 Since the origin of the singularity comes from ordered clusters
 in spatial dimensions, this effect should be absent 
 in the present infinite-ranged models.
 The authors in \cite{GBH1, RY1, RY2, GBH2} 
 called this mechanism as the quantum Griffiths singularity.
 However, the essential mechanism is the same as the classical one.
 The quantum nature is relevant for 
 how this mechanism is related to physical quantities.

 Our finding is that, even for infinite-ranged systems, 
 a similar power-law behavior can be observed 
 for the susceptibility distribution. 
 We cannot attribute this behavior to the ordered clusters
 as in the finite dimensional case.
 The effect must be a purely quantum mechanical one
 which can be expressed by fluctuations in imaginary time.
 Therefore, it is natural to expect that
 ordered clusters are formed in the imaginary time direction.
 Compared to the spatial dimension, 
 the crucial difference is that 
 there is no disorder in this direction.
 Theoretically, the imaginary time correlation arises only when 
 the ensemble average is carried out for the free energy, 
 which implies that the present Griffiths-like effect 
 can be obtained for random quantum systems.
 The randomness induces quantum fluctuations,
 which allows a formation of ordered clusters.
 If we accept this scenario, 
 it is not difficult to obtain a power-law form 
 of the gap distribution function.
 Referred to the classical Griffiths mechanism \cite{Sachdev},
 we expect an exponentially small gap 
 $\Delta\sim\Delta_0 e^{-c\tau/(k+1)}$ in $\tau$ is formed 
 with a very small probability $e^{-c\tau}$, 
 where $\Delta_0$, $k$, and $c$ are some constants.
 Summing over $\tau$, we obtain
\be
 {\rm P}(\Delta)\sim\int d\tau e^{-c\tau}\delta\left(
 \Delta-\Delta_0e^{-\frac{c}{k+1}\tau}\right)\sim \Delta^k.
 \label{qG}
\ee
 Since $0\le \tau\le \beta$, 
 the power law can be obtained only for large-$\beta$.
 This feature is consistent with our expectation that 
 the present effect is a purely quantum one and 
 becomes important at low temperatures. 

 This mechanism is not specific to the present infinite-ranged models
 and the finite dimensional systems can share the same property.
 For the finite dimensional systems, 
 the classical and quantum Griffiths singularities appear at 
 the same time.
 They must be treated in a different way and 
 their contributions to physical quantities 
 are expected to be different.
 It is an important issue to be clarified in future studies.

\section{Conclusions}
\label{conc}

 In conclusion, we have discussed quantum spin-glass phase transitions
 in terms of the energy gap.
 We found for the transverse SK model that 
 the gap distribution shows a power-law behavior near the origin.
 As a result, depending on its exponent, 
 the linear, spin-glass, and nonlinear susceptibilities 
 diverge at different points.
 On the other hand, the transverse REM does not show such behavior.
 Although the gap distribution shows a broad form,
 the state of the system can be described by the ground state 
 and the classical treatment is justified for the REM.
 The difference can be understood by the symmetry of the system.
 The role of symmetry can be more closely studied 
 by generalizing the model (\ref{GSK}) to arbitrary-$p$.
 Details are presented elsewhere.

 The present results are considered 
 a universal feature that can be seen in broad random quantum systems 
 since the mechanism of the quantum fluctuation effects induced
 by disorder is independent of the details of the Hamiltonian.
 Our result for the SK model shows that 
 the self-averaging property does not hold at zero temperature
 and the quantum phase transition 
 is determined by rare configurations of disorder.
 Random quantum systems can be totally different 
 from classical thermodynamic systems.
 In order to reveal such properties, 
 we consider that the present approach is useful and 
 can be a supplement to the dynamical mean field theory 
 as discussed in \cite{MH, Takahashi}. 
 Work in this direction is under progress and 
 we hope that such universal properties are clarified in near future.

\section*{Acknowledgments}

 We are grateful to 
 K. Takeda for useful discussions and comments.
 Y.M. acknowledges support from
 the Japan Society for the Promotion of Science.


\end{document}